\documentclass[twocolumn,preprintnumbers,amsmath,amssymb,floatfix]{revtex4}
\usepackage{graphicx}
\usepackage{dcolumn}
\usepackage{bm}
\graphicspath{{./},{./figs/}}
\usepackage[latin1]{inputenc}

\begin{document}


\title{Substrate effects on the exciton fine structure of black phosphorus quantum dots}

\author{J. S. de Sousa$^1$, M. A. Lino$^{1,2}$, D. R. da Costa$^1$, A. Chaves$^1$, J. M. Pereira Jr.$^1$, G. A. Farias$^1$}
\affiliation{$^1$Universidade Federal do Ceará, Departamento de Física, Caixa Postal 6030, 60455-760 Fortaleza, Ceará, Brazil}
\affiliation{$^2$Universidade Federal do Piauí, Departamento de Física, CEP 64049-550, Teresina, Piauí, Brazil}


\begin{abstract}
We study the size-dependent exciton fine structure in monolayer black phosphorus quantum dots (BPQDs) deposited on different substrates (isolated, Si and SiO$_2$) using a combination of tight-binding method to calculate the single-particle states, and the configuration interaction formalism to determine the excitonic spectrum. We demonstrate that the substrate plays a dramatic role on the excitonic gaps and excitonic spectrum of the QDs. For reasonably high dielectric constants ($\varepsilon_{sub} \sim \varepsilon_{Si} = 11.7 \varepsilon_0$), the excitonic gap can be described by a single power law  $E_X(R) = E_X^{(bulk)} + C/R^{\gamma}$. For low dielectric constants $\varepsilon_{sub} \leq \varepsilon_{SiO_2} = 3.9 \varepsilon_0$, the size dependence of the excitonic gaps requires the sum of two power laws $E_X(R) = E_g^{(bulk)} + A/ R^{n} - B/R^{m}$ to describe both strong and weak quantum confinement regimes, where $A$, $B$, $C$, $\gamma$, $n$, and $m$ are substrate-dependent parameters. We also predict that the exciton lifetimes exhibit a strong temperature dependence, ranging between 2-8 ns (Si substrate) and 3-11 ns (SiO$_2$ substrate) for QDs up 10 nm in size.
\end{abstract}

\maketitle


\section{Introduction}

Black phosphorus (BP) has recently become one of the most attractive two-dimensional materials due to its physical and chemical properties. BP exhibits a highly anisotropic band structure with large direct band gap of the order of 2~eV \cite{ref4,ref5,ref6,ref7,ref8, chaves2016}, high carrier mobilities \cite{xia2014,castellanos2015,rudenko2014,rudenko2015,cakir2014,pereira2015}, nonlinear optical response \cite{lu2015}, magneto-optical Hall effect \cite{tahir2015}, integer quantum Hall effect \cite{li2016}, and thermoelectricity \cite{zhang2014}. All these properties make BP a strong candidate for the development of optical and electronic applications.  

It was recently shown that BP exhibits, depending on the substrate, very large exciton binding energies that can withstand large in-plane electric fields, giving rise to excited excitonic states \cite{chaves2015}. Two recent reports on the optical properties of bulk monolayer BP deposited on quartz and Si substrates, by Zhang \textit{et al.} and Li \textit{et al.}, reported photoluminescence peaks at 1.67 eV and 1.73 eV, respectively \cite{zhang2016, li2017}. The assumption that the difference in the peaks energies of both measurements is  caused by the substrate is the main motivation of this work. 

BP quantum dots (BPQDs) have also been already produced. For example, Sofer \textit{et al.} produced BPQDs with few layers with average size of 15 nm \cite{ref12}. Sun \textit{et al.} synthesized BPQDs as small as $2.6\pm1.8$ nm of diameter and  $1.5\pm0.6$ nm of thickness with a wet exfoliation method \cite{ref13}. Zhang \textit{et al.} also fabricated BPQDs by wet exfoliation, obtaining BPQDs with lateral sizes of $4.9\pm1.6$ nm and thicknesses of $1.9\pm0.9$ nm \cite{ref11}. Xu \textit{ et al.} produced BPQDs with average size of $2.1\pm0.9$ nm in large scale by solvothermal synthesis \cite{ref14}. 

From theoretical point of view, BP and their nanostructures have also been intensively investigated. Rudenko \textit{et al.} developed a tight-binding (TB) parameterization for mono and bilayer BP that has become the basis for the theoretical investigation of several BP structures \cite{rudenko2015,rudenko2014}. Pereira \textit{et al.} derived a continuum model to describe the band structure of BP, departing from the paremeterization of Rudenko \textit{et al.}. They also investigated the Landau levels in the mono and bilayer BP \cite{pereira2015}. de Sousa \textit{et al.} proposed new types of boundary conditions for BP nanoribbons with different edge types to be used in theoretical modelling of BP nanostructures within the continuum model \cite{desousa2016}. Zhang \textit{et al.} investigated the electronics properties of BPQDs with different geometries under the effect of external magnetic fields \cite{zhang2015}. Lino  \textit{et al.} studied the additional energy spectrum of small BPQDs, and demonstrated the feasibility of observing Coulomb blocked effects in BPQDs at room temperature \cite{lino2017}.  Substrate effects on the electronic properties of monolayer BP have been investigated by Mogulkoc \textit{et al.} \cite{mogulkoc2016}. They have reported the broadening of the single-particle gap and renormalization of the effective masses of monolayer BP due to the interaction between carriers in BP and substrate polarons. In particular, the single-particle gap broadening can be of the order of 30 meV for BP deposited on SiO$_2$.

In this work, we calculate the size-dependent excitonic fine structure of monolayer (ML) circular BPQDs using a combination of a TB method to calculate single-particle states, and the configuration interaction (CI) formalism to determine the excitonic fine structure. We aim to understand the effect of different substrates on the exciton fine structure of the BPQDs. This paper is organized as follows. The Theoretical background to calculated single-particle and excitonic states, as well as the optical properties are described in Section \ref{sec:model}. Our results are presented in Section \ref{sec:results}, and discussed in Section \ref{sec:discussion}.

\section{\label{sec:model} Theoretical background}

\begin{figure*}[p]
\begin{center}
\includegraphics[width=.8\textwidth,clip=true]{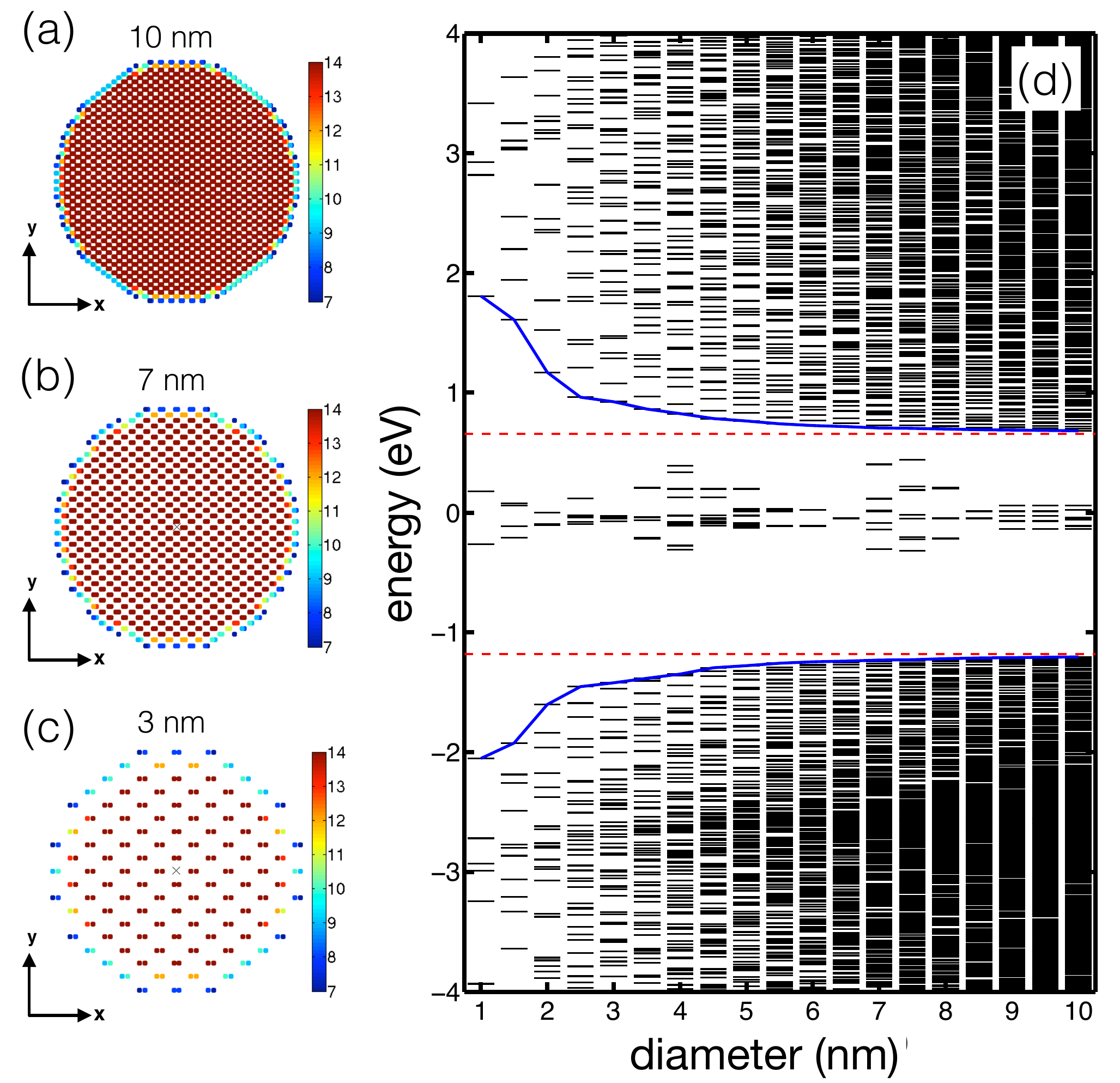}
\caption{\label{fig:BPQDs} Atomic structure of ML-BPQDs with diameters of (a) $10$ nm, (b) $7$ nm and (c) $3$ nm. The center of each structure is marked with a cross. The color schemes indicate the coordination number of each atom assuming a cut-off radius of $0.425$ nm. (d) Size-dependent single-particle energy spectra of isolated BPQDs. Red dashed lines indicate the valence ($e_{vbm}$) and conduction ($e_{cbm}$) band edges of the bulk BP monolayer. Blue solid lines represent the size-dependent band edges of the QDs.}
\end{center}
\end{figure*}

Nearly circular BPQDs were formed by generating a large ML-BP sheet with armchair (zigzag) direction aligned to the $x$ ($y$) axis, and the atoms outside a given radius (measured with respect to center of mass of the large sheet) were disregarded, resulting in QDs with $C_4$ symmetry. Our choice for the dot shape is based on samples produced by exfoliation who exhibits no uniform edges. Some examples of BPQDs studied in this work are depicted in Figures \ref{fig:BPQDs}(a)-\ref{fig:BPQDs}(c). The energy spectrum of the BPQDs was calculated by solving Schrödinger equation represented in a linear combination of atomic orbital (LCAO) basis, such that the effective Hamiltonian reads
\begin{equation}
\hat{H}  = \sum_i \epsilon_i |i\rangle\langle i | + \sum_{i,j} t_{i,j} |i\rangle\langle j |.
\end{equation}
\noindent The generalized index $i = \{\vec{R}_i,\alpha,\nu\}$ represents the orbital $\nu$ of the atomic species $\alpha$ at the atomic site $\vec{R}_i$. $\epsilon_i$ represents the onsite energy of the i-th site, and $t_{i,j}$ represents the hopping parameter between i-th and j-th sites. Since all atoms are identical, the onsite energies only provide an energy reference to the energy spectrum (we adopted $\epsilon_i = 0$ eV). As for the hopping parameters and lattice constants, we adopted the $10$ hopping paramenter TB model of Rudenko \textit{et al.} \cite{rudenko2015}.


\subsection{Excitonic spectra}
The excitons wavefunctions $\Psi_\lambda $ are expressed as linear combination of single substitution Slater determinants $\Phi_{v,c}$ \cite{franceschetti1999,desousa2012}
\begin{equation}
\Psi_\lambda (\vec{r}_e,\vec{r}_h)=\sum_{v}^ {N_v}\sum_{c}^ {N_c}C_{v,c}^{\lambda} \Phi_{v,c},
\end{equation}
where $\lambda$ denotes the exciton quantum numbers, $N_v$ ($N_c$) corresponds to the number of valence (conduction) states included in the expansion. The determinants $\Phi_{v,c}$ are obtained from the ground state (GS) Slater determinant $\Phi_0$ by promoting one electron from the valence band state $\psi_v$ to the conduction band state $\psi_c$:
\begin{eqnarray}
\Phi_0(\vec{r}_1,...,\vec{r}_N)= \mathcal{A}[\psi_1(\vec{r}_1),...,\psi_v(\vec{r}_v),...\psi_N(\vec{r}_N)], \\
\Phi_{v,c}(\vec{r}_1,...,\vec{r}_N)= \mathcal{A}[\psi_1(\vec{r}_1),...,\psi_c(\vec{r}_v),...\psi_N(\vec{r}_N)] ,
\end{eqnarray}
\noindent where $N$ is the number of electrons in the system, and $\mathcal{A}$ is the anti-symmetrisation operator. In the $\{ \Phi_{v,c} \}$ basis, the excitonic spectrum is obtained by solving the following effective Schrödinger equation:
\begin{eqnarray}
\label{eq:excfine}
\sum_{v',c'}^{N_v,N_c}\hspace{-0.2cm}\left[(\epsilon_c\hspace{-0.05cm}-\hspace{-0.05cm}\epsilon_v\hspace{-0.05cm}-\hspace{-0.05cm}E_\lambda)\delta_{v,v'}\delta_{c,c'}\hspace{-0.05cm}-\hspace{-0.05cm}J_{vc,v'c'}\hspace{-0.05cm}+\hspace{-0.05cm}K_{vc,v'c'}\right] C_{v',c'}^{\lambda}\hspace{-0.05cm}=\hspace{-0.05cm}0.\nonumber\\
\end{eqnarray}
\noindent $\epsilon_{c,v}$ represents the single-particle energy states in the conduction and valence bands, respectively.  For simplicity, spin effects in both single-particle and excitonic spectra are reserved for future studies. The quantities $J_{vc,v'c'}$ and $K_{vc,v'c'}$ represent the direct Coulomb and exchange energies:
\begin{eqnarray}
\label{eq:eb} 
J_{vc,v'c'}\hspace{-0.15cm}=\hspace{-0.15cm}\int\hspace{-0.25cm}\int\hspace{-0.15cm}\psi_{v}^{*}(\vec{r}_2)\psi_{c}^{*}(\vec{r}_1)V(|\vec{r}_1\hspace{-0.1cm}-\hspace{-0.1cm}\vec{r}_2 |) \psi_{v'}(\vec{r}_2) \psi_{c'}(\vec{r}_1) d\vec{r}_1 d\vec{r}_2,\nonumber\\
\end{eqnarray}
\begin{eqnarray}
\label{eq:ex} 
K_{vc,v'c'}\hspace{-0.15cm}=\hspace{-0.15cm}\int\hspace{-0.25cm}\int\hspace{-0.15cm}\psi_{v}^{*}(\vec{r}_1)\psi_{c}^{*}(\vec{r}_2)V(|\vec{r}_1\hspace{-0.1cm}-\hspace{-0.1cm}\vec{r}_2 |)\psi_{v}(\vec{r}_1)\psi_{c'}(\vec{r}_2)
 d\vec{r}_1 d\vec{r}_2.\nonumber\\
\end{eqnarray}

\subsection{Screening model}

\noindent The Coulomb interaction potential  $V(|\vec{r}_1-\vec{r}_2 |)$ in two dimensions exhibits a nontrivial form as compared to tri-dimensional bulk materials due to non-local screening effects. We adopted the model of Rodin \textit{et al.} for the Coulomb interaction between charges confined in a two-dimensional material sandwiched between a substrate with dielectric constant $\varepsilon_{sub}$ and vacuum \cite{rodin2014}. This is given by:

\begin{equation}
\label{eq:dielectric}
V(r) = \frac{q^2}{4\pi\varepsilon_0} \frac{\pi}{2 \kappa r_0} \left[ H_0\left( \frac{r}{r_0}\right)-Y_0\left( \frac{r}{r_0}\right)\right],
\end{equation}

\noindent where $r_0 = 2\pi \alpha_{2D}/\kappa$, $\kappa = (1+ \varepsilon_{sub})/2$. $H_0$ and $Y_0$ are the Struve and Neumann functions.  The parameter $\alpha_{2D}  = 4.1$ nm represents the polarizability of a single BP layer in vacuum, and it was determined by Rodin \textit{et al.}, using a density-functional calculations \cite{rodin2014}. 

\subsection{Dipole matrix elements}

The first-order radiative recombination lifetime of the excitonic states $\Psi_\lambda$ is obtained by using the Fermi's golden rule \cite{desousa2012, Leung, Ouisse}
\begin{equation}
\label{eq:rectime} \frac{1}{\tau_\lambda} = \frac{4 n \alpha \omega_{\lambda}^3}{3 c^2} | D_\lambda |^2,
\end{equation}
\noindent where $n = \sqrt{\epsilon_0}$ is the refractive index, $\alpha$ is the fine structure constant, $c$ is the speed of light in vacuum, $\omega_\lambda=E_\lambda / \hbar$, and $D_\lambda$ represents the dipole matrix elements:
\begin{equation}
\label{eq:dipole} D_\lambda = \sum_{v,c} C_{v,c}^{\lambda} \langle \psi_c | \vec{E}_0 \cdot \vec{r} | \psi_v \rangle,
\end{equation}
\noindent where $\vec{E}_0$ is the light polarisation direction. The excitonic absorption cross section can be calculated using Fermi's golden rule as:
\begin{equation}
\label{eq:abs} \sigma(\omega) \propto \sum_{\lambda} |D_{\lambda}|^2 \delta(\hbar \omega - E_{\lambda}).
\end{equation}

\noindent The average exciton lifetime is obtained with:
\begin{equation}
\label{eq:rectime2} \frac{1}{\tau} = \frac{ \sum_{\lambda} \tau_{\lambda}^{-1} e^{-(E_\lambda - E_0)/k_B T}} {\sum_{\lambda}e^{-(E_\lambda - E_0)/k_B
T}},
\end{equation}
\noindent  where $E_0$ is the lowest exciton energy and $k_B$ is the Boltzmann constant.

\section{\label{sec:results}Results}

\subsection{\label{sec:singleparticle}Single-particle spectra}

\begin{figure*}[p]
\begin{center}
\includegraphics[width=.6\textwidth]{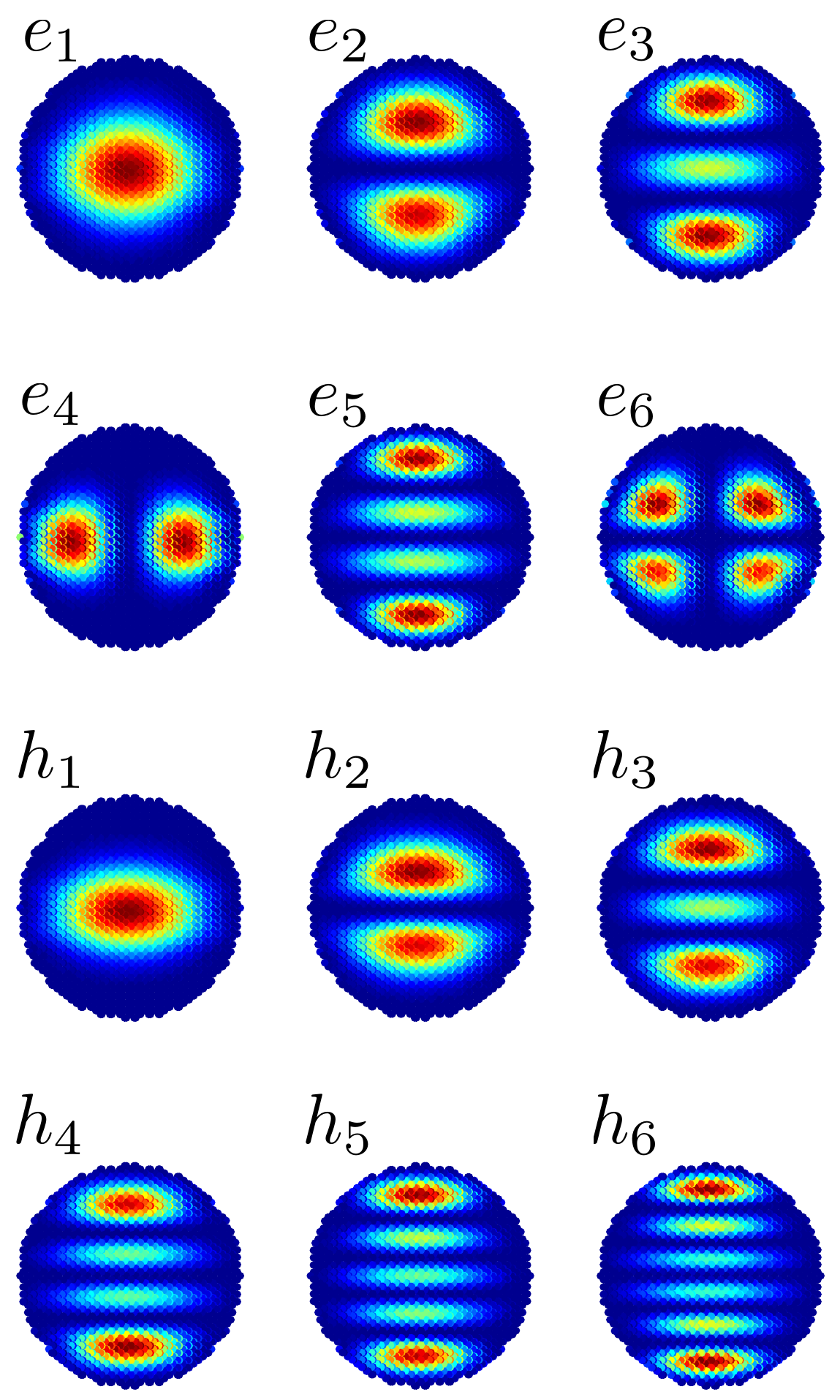}
\caption{\label{fig:wf}Squared wavefunctions of a $10$ nm wide ML-BPQD. Six states in the conduction ($e_n$) and valence ($h_n$) bands are shown.}
\end{center}
\end{figure*}

Figure \ref{fig:BPQDs}(d) shows the size-dependent single-particle spectra of ML-QDs up to 10 nm of diameter.  The horizontal lines represent the conduction $e_{cbm}$ and valence $e_{vbm}$ band edges of the bulk BP monolayer, where $e_{cbm}-e_{vbm} = E_g^{(bulk)} = 1.84$~eV. The blue lines indicate the size-dependent conduction $e_1(R)$ and valence $h_1(R)$ band edges. The size-dependent single-particle bandgap is defined as $E_g(R) = e_1(R) - h_1(R)$, where electron (hole) states are labeled as $e_n$ ($h_m$), and the index $n$ ($m$) grows using as reference $e_{cbm}$ ($e_{vbm}$). There are deep interface states within the bandgap of the QDs and the width of the band of interface states fluctuates with QD size because the QDs are not perfectly circular and exhibit mixed types of edges. Figure \ref{fig:wf} shows the squared wavefunctions of a $10$ nm wide ML-BPQD. The six lowest (highest) confined states in the conduction (valence) band exhibit an increasing number of nodes compatible with two-dimensional quantum confinement with anisotropic effective masses in both conduction and valence bands, whereas the effective masses of electrons and holes in the zigzag ($y$) direction are larger than the ones in the armchair direction. 

The size-dependent single-particle bandgap of isolated ML-QDs is shown in Figure \ref{fig:gap} (black symbols). This quantity can be fitted with the following power law:

\begin{equation}
\label{eq:gap}
E_g(R) = E_g^{(bulk)} + \frac{0.7641}{R^{1.41}},
\end{equation}

\noindent where energies and sizes are in eV and nm units, respectively. Within the effective mass approximation (EMA) framework, it would be expected a size dependence of the type $E_{g} \propto R^{-2}$. The discrepancy between exponents reveals that EMA is not suitable to model the size-dependent bandgap of BPQDs because their actual confinement barrier is not infinite. 

From the single-particle gap, one can perturbatively estimate the excitonic gap as $E_X = E_g - E_B$, where  $E_B=J_{e_1h_1,e_1h_1}$ is the exciton binding energy of the $(e_1,h_1)$ pair. The effect of the substrate is included in the dielectric screening model of Equation (\ref{eq:dielectric}). We have adopted three different substrates: vacuum ($\varepsilon_{vac} = 1$), SiO$_2$ ($\varepsilon_{SiO_2} = 3.9$) and Si ($\varepsilon_{Si} = 11.7$). The size-dependent excitonic gaps (shown in top panel of Figure \ref{fig:gap}) of ML-QDs deposited on Si (red symbols) and SiO$_2$ (magenta symbols) substrates are, respectively, well fitted by the following expressions:

\begin{equation}
\label{eq:exsi_a}
E_X^{(Si)}(R) = 1.69 + \frac{0.6713}{R^{1.41}},
\end{equation}

\begin{equation}
\label{eq:exsio2_a}
E_X^{(SiO_2)}(R) = 1.59 + \frac{0.4415}{R^{1.82}},
\end{equation}

\noindent but the excitonic gap $E_X^{(vac)}(R)$ for isolated ML-QDs  (in vacuum, blue symbols) seems to exhibit two size-dependent regimes and cannot be fitted by a single power law. The size dependence of $E_B$ (bottom panel of Figure \ref{fig:gap}) evidences the strong effect of the substrate on the excitonic gap. In vacuum, $E_B$ varies from $1.1$ eV to $0.47$ eV, when the QD size reduces from $1$ nm to $10$ nm of diameter. For ML-QDs deposited on SiO$_2$ (Si), the binding energies reduces from $0.7$ eV ($0.39$ eV) to $0.25$ eV ($0.11$ eV) for the same size variation. The trend $E_B^{(vac)}>E_B^{(SiO_2)}>E_B^{(Si)}$ is explained by the fact that the electron-hole interaction is inversely proportional to the dielectric constant of the substrate (see Equation (\ref{eq:dielectric})). 

Assuming that the calculations of $E_X(R) = E_g(R)- E_B(R)$ up to $R = 5$ nm are sufficient to capture the bulk behavior when fitting the datasets with $E_X(R) = E_X^{(bulk)} + A/R^n$, we can compare the calculated excitonic gaps with recent photoluminescence (PL) measurements in bulk monolayer BP with the fitted values of $E_X^{(bulk)}$ in Equations (\ref{eq:exsi_a}) and (\ref{eq:exsio2_a}). Zhang \textit{et al.} reported a PL peak at $1.67$ eV for monolayer BP deposited on quartz (same dielectric constant of SiO$_2$) \cite{zhang2016}, and Li \textit{et al.} reported a PL peak at $1.73$ eV for BP deposited on sapphire (same dielectric constant of Si) \cite{li2017}, as shown by the dashed lines in red and magenta in the top panel of Figure \ref{fig:gap}, respectively. If we compare the PL peak of Li (Zhang) at $1.73$ eV ($1.67$ eV) with our fitted  $E_X^{(bulk)}$ value of $1.69$ eV ($1.59$ eV), we obtain a difference of $0.04$ eV ($0.08$ eV) that corresponds to errors of $\approx2.3\%$ ($\approx4.8\%$). This is a strong evidence of the robustness of our TB approach. Thus, the bulk estimates of the exciton binding energies in ML-BP are $E_B^{(Si)} = 0.15$ eV and $E_B^{(SiO_2)} = 0.25$ eV.

When comparing the size-dependence of the exciton interaction and quantum confinement energies (through the ratio $\beta = E_B/E_{conf}$) for the different substrates (see the inset of bottom panel in Figure \ref{fig:gap}), the transition from strong ($\beta<1$) to weak quantum ($\beta>1$) confinement regimes occurs at different sizes, depending on the type of substrate. For ML-BPQDs in vacuum, strong confinement regime only occur for very tiny QDs ($R\leq 1$ nm). As the size of isolated QDs increases, the exciton interaction becomes much stronger than the quantum confinement (eg. $\beta \approx 10$ for $R = 5$ nm). On the other hand, the transition from strong to weak confinement regime occur $R \approx 5$ nm for QDs deposited in Si. Even for large sizes ($R = 5$ nm), the quantum confinement energy is still moderately large compared to the exciton interaction ($\beta \approx 2$). This explain why the exponent of the size-dependence of $E_X^{(Si)}(R)$ is the same of the single-particle gap $E_g(R)$. In the case of SiO$_2$, the relatively low dielectric constant causes the strong-to-weak confinement transition to occur at $R \approx 2$ nm. For large $R$, the exciton interaction quickly becomes dominant ($\beta = 5$, for $R=5$ nm). In this case, the size-dependence exponent $E_X^{(SiO_2)}(R)$ becomes different from the exponent of the size-dependent single-particle gap.

\subsection{Excitonic spectra}

Figure \ref{fig:exc} shows the size-dependent excitonic spectra of BPQDs, where the bandgap interface states were disregarded. Those spectra were calculated using the CI formalism (six states from conduction and valence bands)  considering the possibility of pure (blue lines) and mixed (red lines) exciton configurations. In the mixed configuration, the exciton states are formed by a linear combination of electron-hole $(e_n,h_m)$ pairs, whereas in the pure configuration, only degenerate exciton states are allowed to mix. Furthermore, in the pure configuration, all GS excitons are formed by the single $(e_1,h_1)$ pair, while for the mixed configuration, the composition of the GS exciton is size- and substrate-dependent (see Table \ref{tab:composition}). In the Si substrate, the GS exciton in the QD with $1$ nm of diameter is $99.7\%$ formed by the $(e_1,h_1)$ pair. In the QD with diameter of $10$ nm, the GS exciton is formed by the pairs $(e_1,h_1)$ ($77.6\%$) and $(e_2,h_2)$ ($13.8\%$). In the SiO$_2$ substrate, the exciton composition is more complex due to the enhanced Coulomb interaction (compared to Si substrate) that favor the participation of deeper conduction and valence states even in the GS exciton. For example, in small QDs (up to $2$ nm of diameter), the GS exciton is nearly $100\%$ formed by the $(e_1,h_1)$ pair. The contribution of this pair reduces as the QD size increases, being as low as $50\%$ for QDs of $10$ nm of diameter. 

The mixed CI method lowers the exciton band gap, as shown in the inset panels of Figure \ref{fig:exc}. This reduction in energy depends on the dielectric constant of the substrate, being of the order of $0.02$ eV for Si, and $0.05$ eV for SiO$_2$. If we use the largest QD size ($10$ nm of diameter) as ruler to compare our calculations with the available experiments in bulk BP, this energy reduction makes the excitonic gap of BP on Si to agree even better with the $1.73$ eV PL peak of Li \textit{et al.} \cite{li2017} (sapphire substrate), as compared to the perturbative excitonic gap of Equation (\ref{eq:exsi_a}) (for $R\rightarrow \infty$). On the other hand, the many body interactions included in the CI method improves very little the agreement of Equation (\ref{eq:exsio2_a}) (for $R\rightarrow \infty$) with the $1.67$ eV PL peak of Zhang \textit{et al.} \cite{zhang2016} (quartz substrate). Unfortunately, up to now there are no experimental reports of ML-BP deposited in substrates with dielectric constants lower than $\varepsilon_{SiO_2}$ to compare with our calculations of isolated BPQDs. 

\begin{figure*}[p]
\centerline{\includegraphics[width=.9\textwidth]{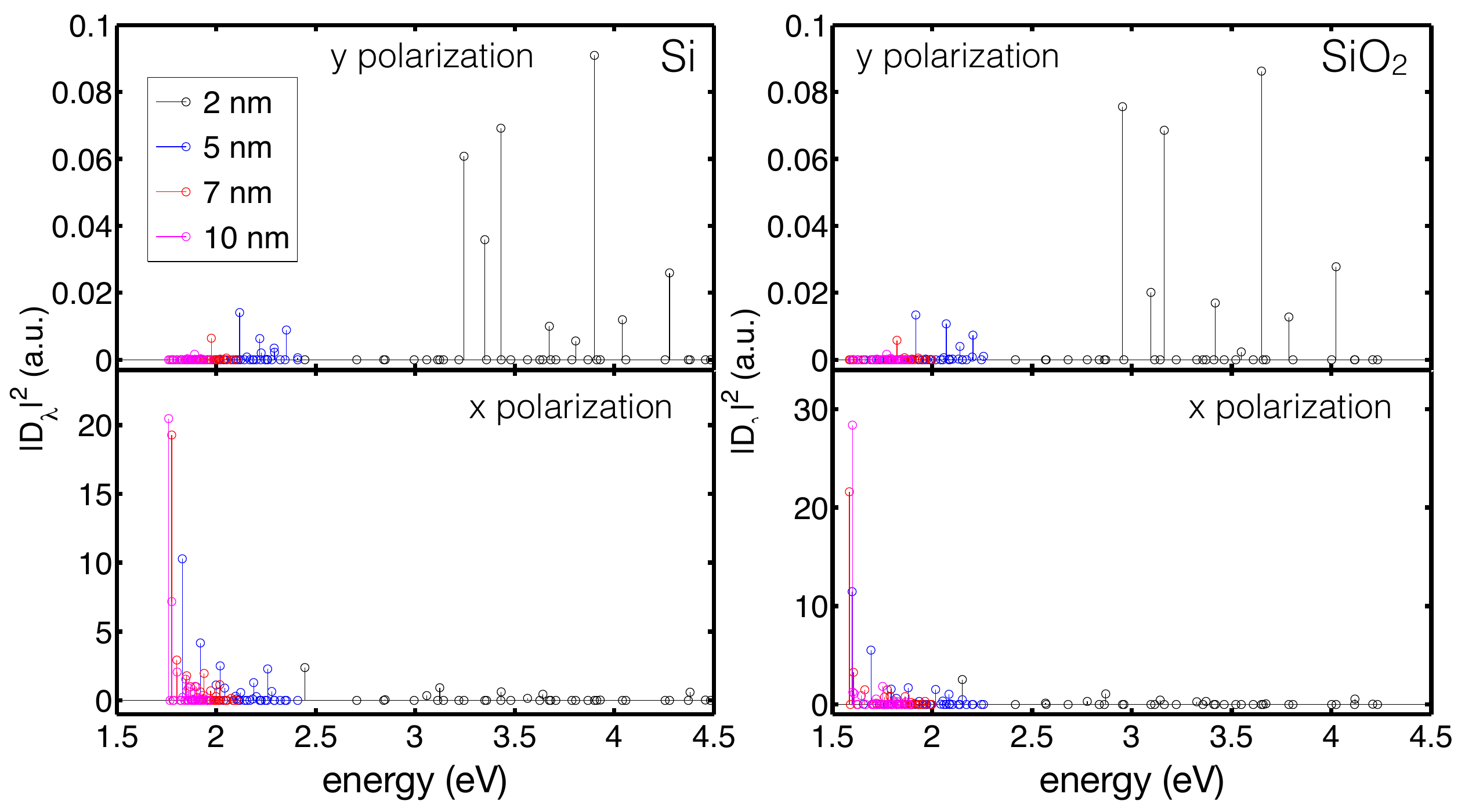}}
\caption{\label{fig:osc}Squared dipole matrix elements $|D_\lambda|^2$ as a function of the exciton energy $E_\lambda$, for light polarization pointing to $y$ direction (top panels) and $x$ direction (bottom panels), and BPQDs deposited on Si (left panels) and SiO$_2$ (right panels) substrates.}
\end{figure*}

BPQDs display a rich and complex size-dependent excitonic structure, exhibiting dark and bright exciton states, where only bright excitons contribute to light-emitting processes. The squared dipole matrix elements $|D_\lambda|^2$ are shown in Figure \ref{fig:osc}. A strong anisotropy associated to the orientation of light polarization is observed. The light polarization pointing to $x$ direction (parallel to the armchair direction) results in matrix elements $2$ orders of magnitude larger than for the light polarization in $y$ direction (parallel to zigzag direction). This is compatible with recent PL experiments in bulk BP, which demonstrated that its PL emission has no optical signal in $y$ direction \cite{li2017}. Besides that, the polarization in $x$ direction (bottom panels) exhibits strong optical activity in the lower part of the excitonic spectra, while the polarization in $y$ direction (top panels) exhibits weak optical activity at higher energies. Experimental absorption peaks are proportional to $|D_\lambda|^2$. Thus it is expected that the absorption peaks increase either with the QD size or with a reduction of the substrate dielectric constant.

\begin{figure*}[p]
\begin{center}
\includegraphics[width=.75\textwidth]{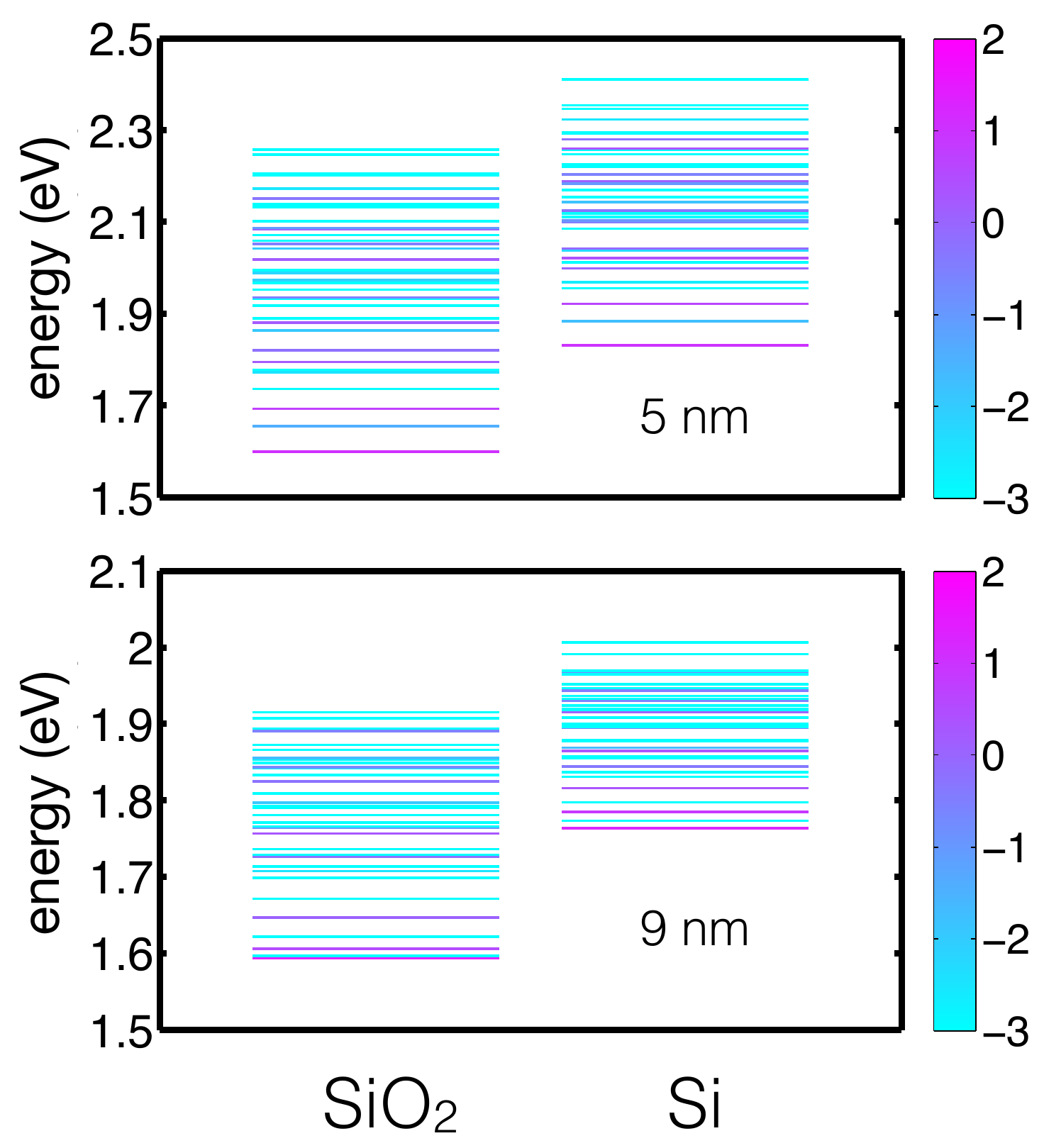}
\caption{\label{fig:exc_compare}Comparison of the fines structure of excitons in BPQDs with $5$ nm (top panel) and $9$ nm (bottom panel) of diameter in different substrates. The color bars indicate the $log|D_\lambda|^2$  calculated for x polarization of each $E_\lambda$ excitonic state.}
\end{center}
\end{figure*}

In Figure \ref{fig:exc_compare} we compare the fine structure of excitons in different substrates. For the case of the BPQD with $5$ nm of diameter (top panel), the GS exciton is the brightest one in both substrates. The first excited exciton states is dark and located $56$ meV and $52$ meV above the GS for SiO$_2$ and Si substrates, respectively. The next bright excitons are $94$ meV ($48\%$ weaker than the GS) and $90$ meV ($59\%$ weaker than the GS) above GS for SiO$_2$ and Si substrates, respectively. The BPQD with $9$ nm of diameter on SiO$_2$ exhibits almost doubly degenerated GS exciton ($\Delta E\approx 3$ meV), where the GS is bright and the first excited state is dark ($|D_0|^2 << |D_1|^2 \approx 10^1$). The next bright state is $13$ meV above (88\% weaker than) the GS. For the BPQD on Si with $9$ nm of diameter, the separation between the two lowest bright states is $22$ meV, with the second bright exciton being 58\% weaker than the GS. 

\begin{figure*}[p]
\begin{center}
\includegraphics[width=.75\textwidth]{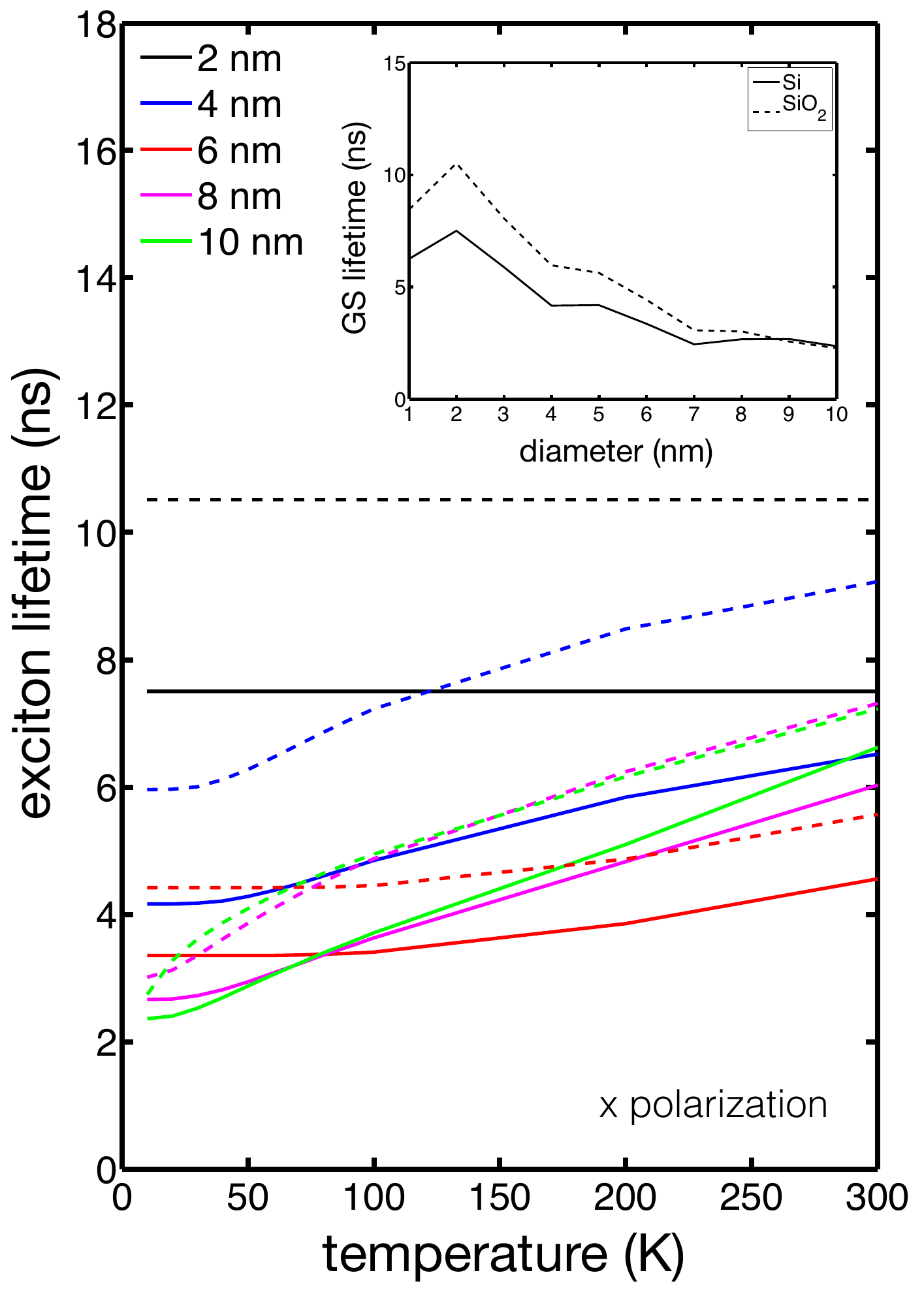}
\caption{\label{fig:lifetime}Temperature dependence of the average excitons lifetime in BPQDs. Lifetimes for $y$ polarization (not shown here) are many orders of magnitudes larger than for $x$ polarization. Results for Si and SiO$_2$ substrates are represented by solid and dashed lines, respectively.}
\end{center}
\end{figure*}

It is instructive to investigate the temperature dependence of the average excitons lifetime for light polarization in $x$ direction, shown in Figure \ref{fig:lifetime}. At low temperatures, the exciton lifetime is inversely proportional to the BPQD size, while at room temperature this relationship becomes more complicated, probably because of changes in the QD interface as the size of the BPQDs grows, affecting the energy distribution and wavefunctions of excited single-particle states, and consequently, excitonic states. The low temperature dependence is in general dominated by the lifetime of the ground state exciton, which also exhibits an inversely proportional relationship with QD size (shown in the inset for Si (solid lines) and SiO$_2$ (dashed lines) substrates). The lifetime of excitons in small BPQDs are insensitive to temperature (see black curves in Figure \ref{fig:lifetime} for dots with $2$ nm of diameter, respectively), while for larger BPQDs, the exciton lifetimes exhibit a monotonic increase with temperature. The substrate has a dramatic effect: the average exciton lifetime is inversely proportional to $\varepsilon_{sub}$.   The exciton lifetimes for light polarization in $y$ direction are not shown here because they are many orders of magnitude larger than the lifetimes for $x$ polarization. 

\section{\label{sec:discussion}Discussion and conclusions}

BPQDs exhibit a size-dependence single-particle bandgap $E_g(R) - E^{bulk}_{g} \propto R^{-1.41}$ in disagreement with simple models based on the EMA, where an exponent $n = 2$ is expected even for two-dimensional QDs. For example, Si nanocrystals have been intensively investigated in the nineties by different atomistic methods \cite{lwwang,hill1995,ogut1997,rohlfing1998,delerue2000,vasiliev2001}, and those studies also found exponents $n < 2$ for the size-dependent Si nanocrystal bandgaps. This discrepancy between EMA and atomistic theories to explain size-dependence of the bandgap of quantum dots is well known. The exponent $n = 2$ arises from infinite confinement barriers (vanishing wavefunctions) to simplify boundary conditions. Exponents with $n < 2$ can be obtained if one considers finite confinement barriers. However, the exponent $n = 1.41$ seems to be related to mixture of border geometries in our circular BPQDs. This conclusion is based on the recent theoretical study of de Sousa \textit{et al.} \cite{desousa2016} showing that the band gap of zigzag and armchair BP nanoribbons scales with $1/D$ and $1/D^2$ ($D$ is the width of the nanoribbon), respectively. Our exponent $n = 1.41$ is in qualitative agreement with the fact that BPQDs with mixed borders should exhibit an intermediate exponent between 1 and 2. 

When excitonic effects are taken into account, the bandgap strongly depends on the substrate. For Si and SiO$_2$, the excitonic gap (calculated perturbatively using a single-particle approach) obeys a single power law $E_X(R) - E^{bulk}_{g} \propto R^{-n}$, where the exponent $n$ is substrate-dependent. For isolated QDs (vacuum as substrate), the size-dependent excitonic gap seems to obey a combination of power laws to describe two different regimes of strong (small QDs) and weak (large QDs) quantum confinement. One can generalize the size dependence of the excitonic band gap of QDs with the simple expression
\begin{equation}
\label{eq:gapgeneral}
E_X(R) = E_g^{(bulk)} + \frac{A}{R^m} - \frac{B}{R^n},
\end{equation}
\noindent where the second and third terms represent the power laws describing the quantum confinement and exciton binding energies, respectively. The parameters $A$, $B$, $m$ and $n$ depend on several factors like dimensionality of the quantum confinement, surface passivation, effectives masses, and dielectric mismatch between QD and the external materials. In the case of our unpassivated BPQDs, $E_g^{(bulk)}$, $A$ and $m$ are known (see Equation (\ref{eq:gap})). However, some phenomenological assumptions can be made. For example, it is known that: (i) $m \leq 2$ ($m=2$ for infinite confinement barriers within EMA); (ii) $n\leq 1$; and (iii) $m>n$. It also known that $B^{-1} \propto \Gamma(\varepsilon_{in},\varepsilon_{out})$, where $\Gamma(\varepsilon_{in},\varepsilon_{out})$ represents a relationship describing the dielectric mismatch. 

\begin{figure*}[p]
\begin{center}
\includegraphics[width=.7\textwidth]{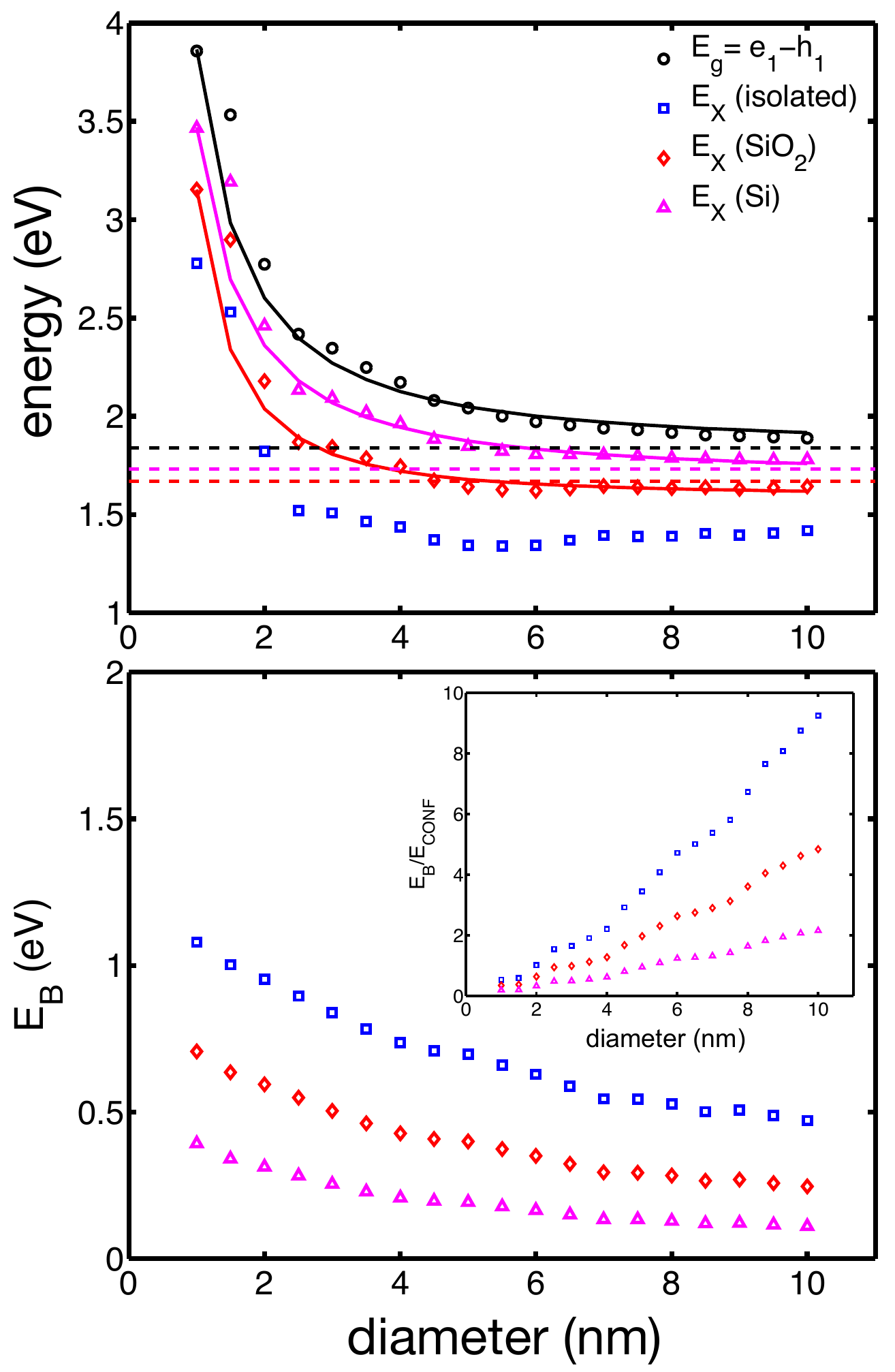}
\caption{\label{fig:gap}(top) Symbols represent the size-dependent single-particle bandgap $E_g$ and excitonic gaps $E_X$ in different substrates. Solid lines are fitting expressions. Dashed lines represent the bulk ML-BP single-particle bandgap (black), PL peak of the bulk monolayer deposited on sapphire (same dielectric constant of Si) measured by Li \textit{et al.} \cite{li2017} (magenta), and PL peak of the bulk monolayer deposited on quartz (same dielectric constant of SiO$_2$) measured by Zhang \textit{et al.} \cite{zhang2016}. (bottom) Size dependence of the fundamental exciton binding energy $E_B$ in different substrates. The inset graph shows the ratio between $E_B$ and the quantum confinement energy $E_{conf} = E_g(R) - E_g^{(bulk)}$.}
\end{center}
\end{figure*}

\begin{figure*}[p]
\begin{center}
\includegraphics[width=.75\textwidth]{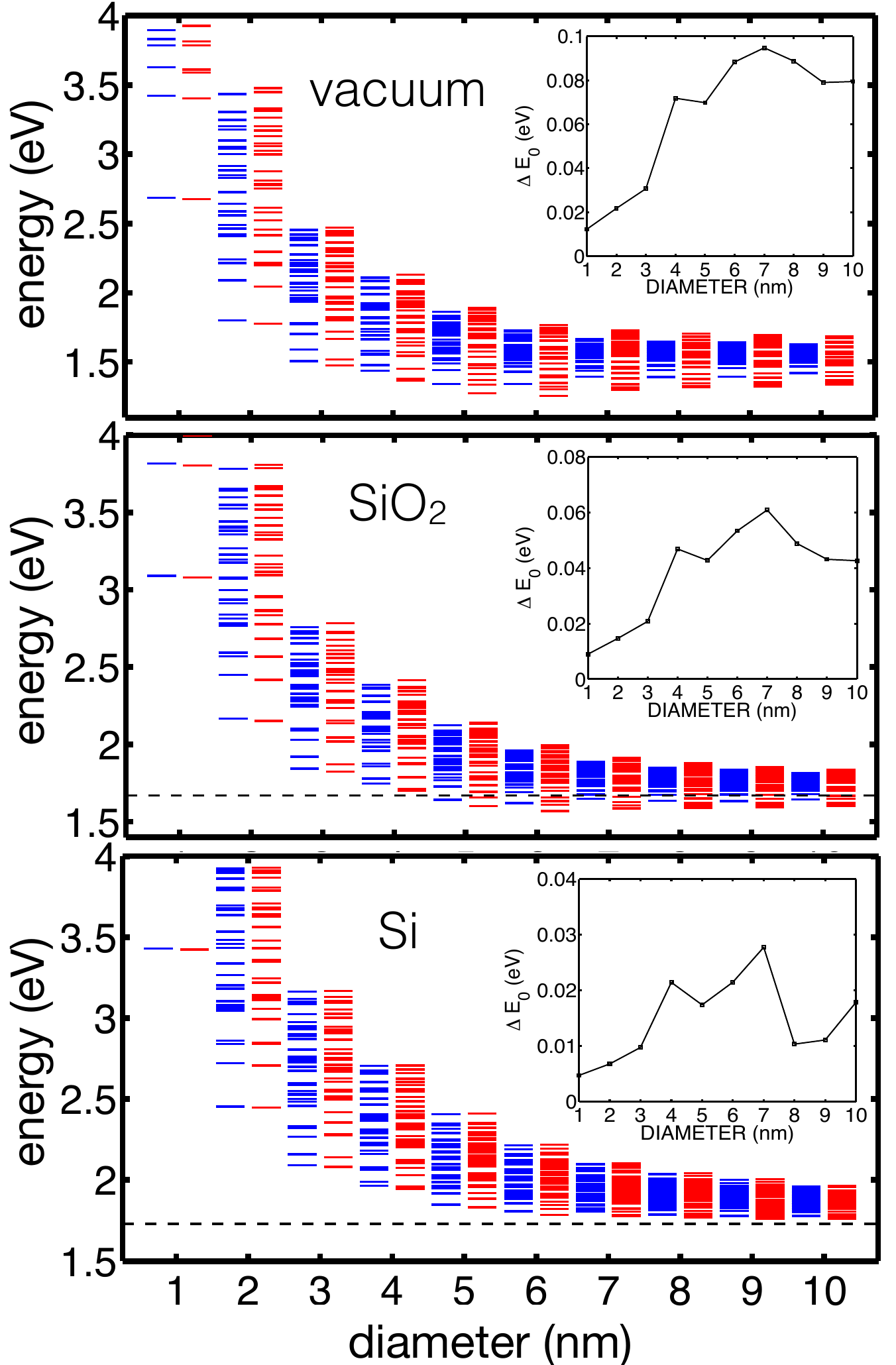}
\caption{\label{fig:exc}Size-dependent excitonic spectra calculated with CI formalism (using six states from each conduction and valence bands) considering pure (blue lines) and mixed (red lines) configurations. Results for isolated BPQDs, as well as QDs deposited on SiO$_2$ and Si are shown in top, middle and bottom panels, respectively. The inset graphs show the difference between GS excitons calculated without and with mixed configurations. The GS excitons composition in this figure are listed in Table \ref{tab:composition}. The black dashed lines in the middle and bottom panels represent the PL peaks measured in a monolayer BP by Zhang \textit{et al.} ($1.67$ eV) and Li \textit{et al.} ($1.73$ eV), respectively \cite{zhang2016, li2017}.}
\end{center}
\end{figure*}

\begin{figure*}[p]
\begin{center}
\includegraphics[width=.85\textwidth]{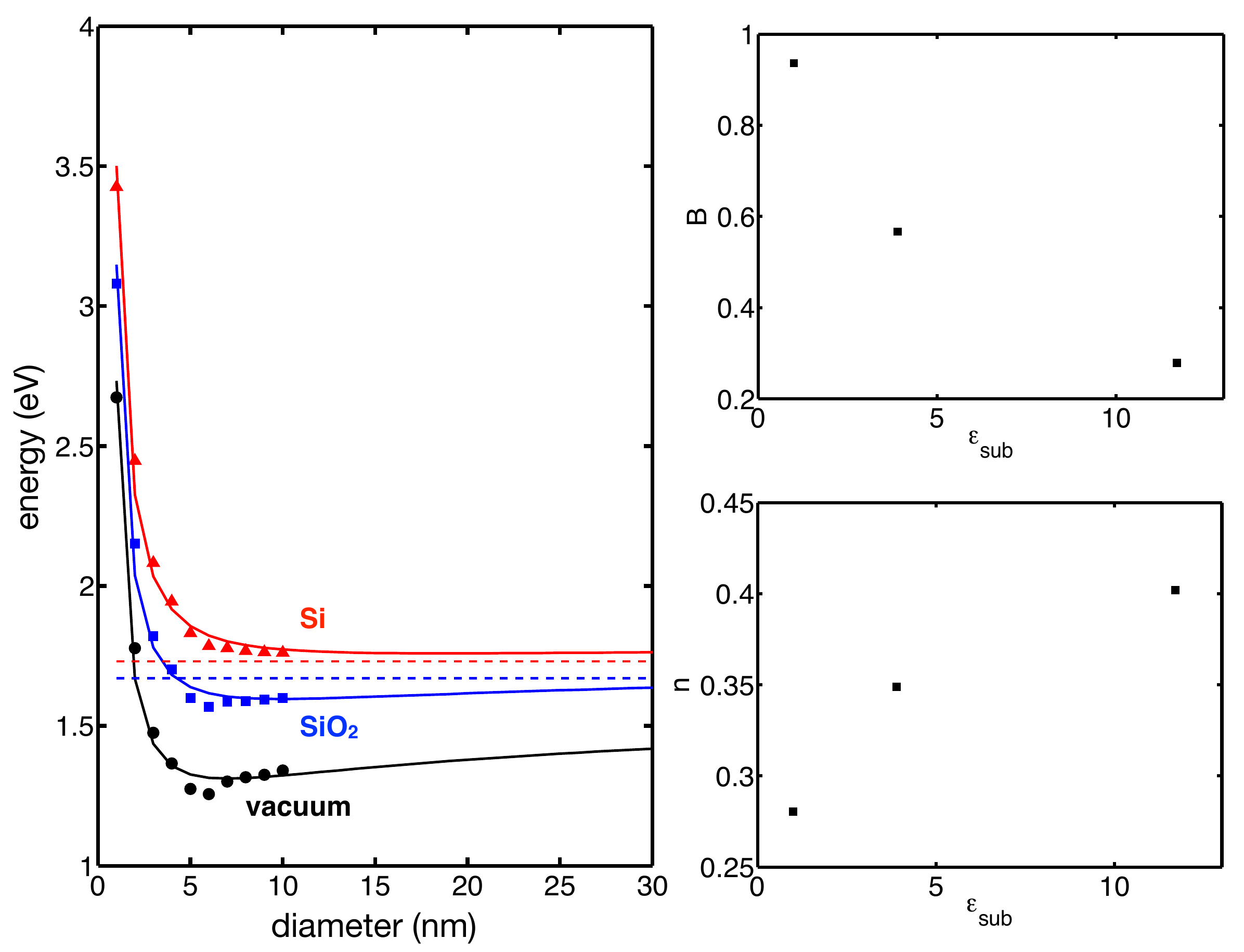}
\caption{\label{fig:gapgeneral}Analysis of size and substrate dependence of the excitonic gaps with Equation (\ref{eq:gapgeneral}). Symbols represent the gaps calculated with CI method, and lines are the fitted curves. Parameters $A$ and $m$ are given in Equation (\ref{eq:gap}). The right panels depict the fitting parameters $B$ and $n$ as a function of $\varepsilon_{sub}$. The dashed lines in the left panel represent the PL peaks measured in a monolayer BP by  Zhang \textit{et al.} (blue, $1.67$ eV) and Li \textit{et al.} (red, $1.73$ eV), respectively \cite{zhang2016, li2017}.}
\end{center}
\end{figure*}

In the most general form of Equation (\ref{eq:gapgeneral}), the two size regimes are separated by a minimum point (see the vacuum case in Figure \ref{fig:gap}, and the vacuum and SiO$_2$ cases in Figure \ref{fig:exc}). The appearance of this minimum point is unexpected compared to the single power law observed both theoretical and experimentally in many types of quantum confined structures. This minimum point $R_{min}$ is located at $R_{min}^{m-n} = (m/n)(A/B)$, and the double power law behavior disappears when $R_{min}\rightarrow \infty$. In the case of BP, we have $B \propto \varepsilon_{sub}^{-1}$. Therefore, the position of the minimum point is directly proportional to $\varepsilon_{sub}$ in qualitative agreement with Figure \ref{fig:exc}. In Figure \ref{fig:gapgeneral}, we fit the excitonic gaps calculated with the CI method with Equation (\ref{eq:gapgeneral}). We obtain that the parameter $B$ ($n$) is inversely (directly) proportional to $\varepsilon_{sub}$.


The perturbative approach adopted in Section \ref{sec:singleparticle} allows to estimate the ground state exciton of 1.59 and 1.69 for ML-BP deposited on SiO$_2$ and Si, respectively. Those results exhibit a remarkable agreement when respectively, compared to the measurements of Zhang (1.67 eV, quartz substrate) and Li (1.73 eV, sapphire substrate) \cite{zhang2016, li2017}. The errors between theory and experiments are 0.08 eV for SiO$_2$ and 0.04 eV for Si substrates. The estimated bulk exciton binding energies are $E_B^{(SiO_2)} = 0.25$ eV and $E_B^{(Si)} = 0.15$. Zhang \textit{et al.} used a simple TB model to explain their measurements \cite{zhang2016} (their bulk bandgap was $2.12$ eV), resulting in an exciton binding energy of $0.45$ eV (quartz substrate), which is $80\%$ larger than our estimate using SiO$_2$ as substrate ($\varepsilon_{quartz}\approx \varepsilon_{SiO_2}$). Li \textit{et al.} explained their measurements with a simple TB model \cite{li2017}, but using a bulk bandgap of $1.8$ eV, leading to a binding energy of $0.07$ meV for monolayer BP on Si substrate. Here, our estimated binding energy is $50\%$ larger than the value of Li \textit{et al.}. Despite of those discrepancies, our method is in very good quantitative and qualitative agreement with those state-of-the-art measurements. The actual values of single-particle gaps and exciton binding energies are still under debate. Several theoretical and experimental reports in the literature use bulk bandgaps varying between $1.52$ eV and $2.12$ eV \cite{rudenko2014,rudenko2015, zhang2016, li2017}, and accurate values are necessary in order to determine actual values of exciton binding energies. We believe that the ten-parameter TB scheme of Rudenko \textit{et al.} is, so far, the most accurate band structure description of BP in the literature \cite{rudenko2015}. In addition, another critical issue is the understanding of the role of dielectric screening in two-dimensional materials \cite{castellanos2015,rodin2014,cartoixa2005,cudazzo2011,berkelbach2013,latini2015,olsen2016}. 

The inclusion of many-body effects within the CI framework allows us to calculate a number of features which cannot be predicted by simple single-particle methods. We have calculated the excitonic spectra for BPQDs on different substrates and their optical properties. Several experimental studies reported an extraordinary dependence of the optical properties of BP with respect to the direction of light polarization with a rich set of optical resonances appearing for light polarisation in the armchair direction \cite{zhang2016,li2017,zhang2014b}. For example Li \textit{et al}. reported strong PL and absorption signals polarized in $x$ direction (armchair direction) and no signal at all with $y$ polarisation (zigzag direction). This is in good qualitative agreement with the ratio of $10^2$ between the calculated squared dipole matrix elements for polarisation in $x$ and $y$ directions. Finally, Zhang \textit{et al}. reported strong temperature dependence of the Raman phonon modes in few layer BP, which is consistent with the strong temperature dependence of the excitonic lifetimes of BPQDs with diameter larger than $4$ nm \cite{zhang2014b}.

The single-particle perturbative approach provided good estimates to the excitonic gaps determined by the CI method. The agreement is particularly good for Si substrate. For SiO$_2$ substrate exhibits some discrepancies for diameter ranging between $3$ nm and $7$ nm. The size-dependent excitonic gaps of the CI method clearly exhibit a shape that resembles a sum of power laws, as in the case of isolated QDs (See Figure \ref{fig:gap}). Using Equation (\ref{eq:gapgeneral}) to fit all excitonic gaps calculated with CI method and extrapolating the results for very large sizes (see Figure \ref{fig:gapgeneral}), the fitted gaps seem to converge to values very close to the measurements of Zhang and Li \cite{zhang2016,li2017}. It is remarkable that QDs as large as $R = 5$ nm are still far from monolayer bulk behavior when deposited in substrates with very low dielectric constants. We remark that subtle effects like the coupling of charges in BP and substrate polarons induces broadening of the single-particle gap and renormalization of effective masses (specially on zigzag direction). For example, Mogulkoc have shown the single-particle gap of ML-BP deposited on SiO$_2$ are enlarged by 30 meV \cite{mogulkoc2016}. If such effects were included in our model, the agreement of our calculations using SiO$_2$ as substrate with the experimental results of Zhang \cite{zhang2016} would be even better. Another possible ingredient to improve the quantitative agreement with experimental measurements is the increase of the CI basis size with more than six electron and hole states, because the enhanced Coulomb interaction in substrates with low dielectric constants may mix even deep electron-hole pairs.

Despite of the good agreement of our calculations if the limit of large BPQDs with experimental measurements in ML-BP, one might argue that our model does not take into account complicated edges effects. Liang \textit{et al.} studied edges reconstruction in ML-BP combining scanning tunelling spectroscopy (STS) and theoretical methods based on the Density Functional Theory (DFT) \cite{liang2014}. They reported that most dangling bonds self-passivate such that the coordination number of phophorus increase from 3 (in the middle of BP layer) to 4 or 5 at the edges, depending on the type of edge geometry. They calculated the electrostatic potential in zigzag BP nanoribbons to account for local fields near the edges due to reconstruction of dangling bonds. They show that the edge reconstruction creates a localized short-range ($\approx$ 0.15 nm) confining potential of 0.15 eV at the edges of a BP layer in vacuum. For supported BP layers, as shown in our calculations, this local edge files would be inversely proportional to the dielectric constant of the substrate. Even for dielectric constants as low as the one of SiO$_2$, the edges contribution would represent a small perturbation compared to the actual size-dependent single-particle band gap of small BPQDs. For larger BPQDs, their effects should be negligible. On the other hand, passivation of the dangling bonds with other atomic species like hydrogen and oxygen due to the exposition of BPQDs to air are expected to eliminate interface states and lower band gaps \cite{garoufalis2006}. Anyhow, a clear picture of the effects of edge reconstruction and/or passivation in the excitonic properties of BPQDs is an open question that must be further investigated.

In conclusion, we studied the excitonic interactions in ML-BPQDs with a realistic TB scheme to calculate single-particle states and the CI method to account for many-body effects. These combination of methods allowed us to (i) reproduce well the results of state-of-the-art experiments of a couple of  groups using substrates with different dielectric constants ranging from reasonably strong (SiO$_2$) to weak (Si) dielectric screening, and to (ii) predict excitonic properties of BPQDs on different substrates. Despite of the success in the synthesis of small BPQDs, the fine excitonic structure of BPQDs have not yet been reported, and the predictions made in this works have yet to be confirmed.
\\

\noindent\textbf{Acknowledgements} The authors acknowledge the financial support from the Brazilian National Research Council (CNPq) and CAPES foundation.
\\


\begin{table*}[t]
\caption{\label{tab:composition}Composition of the GS excitons of Figure \ref{fig:exc}. Components weighting less than $5\%$ are not listed. }
\begin{center}
\small
\begin{tabular}{cl} \hline\hline
diameter & GS exciton composition \\ \hline \hline
\multicolumn{2}{c}{vacuum} \\ \hline 
1 nm & $(e_1,h_1)$: 99.3\% \\ \hline
2 nm & $(e_1,h_1)$: 97.6\% \\ \hline
3 nm & $(e_1,h_1)$: 92.2\% , $(e_2,h_2)$: 5.0\%  \\ \hline
4 nm & $(e_1,h_1)$: 56.5\% , $(e_1,h_2)$: 16.2\% , $(e_2,h_1)$: 11.5\%, $(e_2,h_2)$: 8.7\%  \\ \hline
5 nm & $(e_1,h_1)$: 81.9\% , $(e_4,h_6)$: 6.6\% \\ \hline
6 nm & $(e_1,h_1)$: 74.6\% , $(e_3,h_1)$: 5.0\% , $(e_4,h_6)$: 9.8\% \\ \hline
7 nm & $(e_1,h_1)$: 60.7\% , $(e_2,h_2)$: 22.6\% , $(e_3,h_3)$: 8.2\% \\ \hline
8 nm & $(e_1,h_2)$: 33.7\% , $(e_2,h_1)$: 19.6\% , $(e_2,h_3)$: 22.1\% , $(e_3,h_2)$: 8.4\% , $(e_3,h_4)$: 10.2\% \\ \hline
9 nm & $(e_1,h_2)$: 26.1\% , $(e_2,h_1)$: 12.0\% , $(e_2,h_3)$: 30.2\%,  $(e_3,h_2)$: 11.5\%, $(e_3,h_4)$: 12.9\% \\  \hline
10 nm & $(e_1,h_2)$: 19.4\% , $(e_2,h_1)$: 10.0\% , $(e_2,h_3)$: 28.8\%,  $(e_3,h_2)$: 10.6\%, $(e_3,h_4)$: 17.1\% , $(e_5,h_3)$: 5.1\% \\  \hline
\multicolumn{2}{c}{SiO$_2$ substrate} \\ \hline 
1 nm & $(e_1,h_1)$: 99.5\% \\ \hline
2 nm & $(e_1,h_1)$: 98.4\% \\ \hline
3 nm & $(e_1,h_1)$: 94.1\%  \\ \hline
4 nm & $(e_1,h_1)$: 76.4\% , $(e_2,h_2)$: 10.7\% \\ \hline
5 nm & $(e_1,h_1)$: 86.9\% \\ \hline
6 nm & $(e_1,h_1)$: 79.9\% , $(e_4,h_6)$: 6.5\%\\ \hline
7 nm & $(e_1,h_1)$: 65.4\% , $(e_2,h_2)$: 21.3\% , $(e_3,h_3)$: 6.2\% \\ \hline
8 nm & $(e_1,h_1)$: 70.2\% , $(e_2,h_2)$: 18.9\% , $(e_3,h_3)$: 5.3\%  \\ \hline
9 nm & $(e_1,h_1)$: 64.5\% , $(e_2,h_2)$: 23.4\% , $(e_3,h_3)$: 7.2\%  \\  \hline
10 nm & $(e_1,h_1)$: 50.0\% , $(e_2,h_2)$: 28.3\% , $(e_3,h_3)$: 11.2\% \\  \hline
\multicolumn{2}{c}{Si substrate} \\ \hline 
1 nm & $(e_1,h_1)$: 99.7\% \\ \hline
2 nm & $(e_1,h_1)$: 99.2\% \\ \hline
3 nm & $(e_1,h_1)$: 96.8\% \\ \hline
4 nm & $(e_1,h_1)$: 87.7\%, $(e_2,h_2)$: 7.8\% \\ \hline
5 nm & $(e_1,h_1)$: 92.8\% \\ \hline
6 nm & $(e_1,h_1)$: 88.3\%\\ \hline
7 nm & $(e_1,h_1)$: 77.9\%, $(e_2,h_2)$: 17.5\% \\ \hline
8 nm & $(e_1,h_1)$: 80.0\%, $(e_2,h_2)$: 11.7\%\\ \hline
9 nm & $(e_1,h_1)$: 82.6\%, $(e_2,h_2)$: 10.1\% \\ \hline
10 nm & $(e_1,h_1)$: 77.6\%, $(e_2,h_2)$: 13.8\% \\ \hline \hline
\end{tabular}
\end{center}
\end{table*}


\end{document}